# Role of Broken Gauge Symmetry in Transport Phenomena Involving Neutral and Charged Particles in Finite Lattices

Scott R. Chubb
*Remote Sensing Division*
*Naval Research Laboratory, Washington, DC 20375, USA*

As opposed to the conventional, approximate theory of electrical conduction in solids, which is based on energy band, "quasi-particle" states in infinite lattices, a rigorous theory exists that can be used to explain transport phenomena, in finite lattices, at reduced temperature, through the effects of a broken gauge symmetry: The loss of translational invariance with respect to Galilean transformations that maintain particle-particle separation. Implications of this result in areas related to the transport of hydrogen (H) and its isotopes in nano-crystalline structures of palladium (Pd) and of neutral, coherent atomic waves in finite, optical lattices are presented. These include: 1. The prediction of a novel variant of a known, phenomenon, Zener/Electronic Breakdown in insulators, in which ions in nano-scale palladium-deuteride (PdD) crystals (as opposed to electrons in insulating crystals) that initially, effectively, are confined to particular regions of space, begin to move, spread-out, and conduct charge after they are subjected to an applied, external (constant) electric field $\vec{E}$ for a sufficiently long period of time; and 2. A rigorous treatment of scattering at low temperatures that can be used to identify critical time- and length- scales for problems related to the transport of neutral atomic matter waves in finite, optical lattices, in the presence of gravitational fields.

## I Introduction

This paper formalizes and generalizes a number of intuitive ideas associated with electron band theory in solid state physics, in the idealized limit of infinitely-repeating, periodic lattices, to situations involving finite structures, with real boundaries. The arguments are general: they can be applied to all ordered or partially ordered solids, in the limit of low, but finite temperature T. In particular, in this paper, the terms order and periodic order are used inter-changeably. A solid that possesses partial order (or partial, periodic order) refers to the requirement that the particles in some finite volume, within the solid, occupy one or more many-body states $\Psi$ possessing some degree of periodic symmetry, for some finite period of time. This last condition is defined by the requirement that one or more Bragg diffraction peaks (Ashcroft & Mermin 1976A), beyond the zeroeth order peak, be observable in the diffraction pattern that results when particular particles (x-rays, neutrons, etc. ) scatter elastically off of the solid. An ordered, periodic, or periodically ordered solid refers to the limit in which the resolution between different diffraction peaks and the number of peaks asymptotically approach the resolution and number that one would conventionally associate with a diffraction pattern involving a situation in which a macroscopic number of scattering centers can be involved with the scattering processes that are responsible for the pattern. For the purpose of determining critical time- and length- scales associated with experiments involving finite size optical lattices, a number of the arguments involving finite size effects (and the associated overlap features) in partially ordered solids can also be generalized to optical lattices. (This topic is addressed in the final section of the paper.)

The underlying logic behind the analysis follows from a number of general results that are required in order to insure that the reaction rate be minimized (through minimal overlap with potential perturbations) for a potential process that can couple the ground state (GS) to the lowest-lying excited states, in a solid or lattice. In particular, in general, the GS of a many-body system is required to have minimal overlap with external perturbations and with the lowest-lying excited states. But in the idealized limit in which a set of measurements can be performed, in such a way that the measurement process does not affect the energy of any of a number of potential many-body states (including the GS), locally, in a particular region of space, each measurement can be viewed as a form of "symmetry operation". When each of these symmetry operations can be related to the others through one or more continuously-varying parameters (for example, through measurements of a continuously-varying angle or displacement, performed outside the particular region of space), the set of symmetry operations is referred to as a gauge symmetry.

But because no symmetry operation is perfect, residual perturbations, tied to the measurement process exist. In the presence of these perturbations, the overlap of the GS with low-lying excited states is minimized through processes that minimize the GS energy. The resulting changes in wave function overlap and energy occur through a process that is referred to as broken gauge symmetry. Provided the particular region (which will be referred to as the bulk region) occurs within the solid (or within a lattice), where the numbers of neutral and charged particles are conserved (relative to a particular, well-defined boundary or boundary region), the rate of change of any process involving the GS and its overlap with low-lying excited states (through residual perturbations) is dominated by a universal broken gauge symmetry that occurs in ordered, finite lattices (the loss of translation symmetry at the boundaries of a lattice). In this particular limit: 1. In the absence of the residual perturbations, the energy remains invariant with respect to the gauge symmetry operation associated with performing a rigid shift of the bulk region, in which the coordinate of each particle in the bulk is shifted by the same, constant amount; 2. Since, by construction, the associated coordinate transformation (which is referred to as a Galilean, coordinate transformation) preserves the separation between each particle and the remaining particles in the bulk, a large degeneracy exists, and (in the presence of broken gauge symmetry) the lowest-lying excited states are all related to the GS through one of the possible (Galilean) transformations; and 3. In the presence of residual interactions, the lowest-lying excited states conserve particle number within the bulk region.

Since the associated transformations do not alter the relative separations between any of the particles, effectively, the resulting change in position of the particles can be viewed as occurring through a form of "perfect" rigid body motion, in which the entire collection of particles moves with a common velocity, that is similar to the nearly-perfect, lattice recoil that occurs when a gamma ray strikes a nucleus, without exciting a lattice vibration, in the Mossbauer effect. In this limit, the overlap between the GS and lowest-lying excited states can involve processes that do not alter the relative separations between particles in the bulk. The associated effects, by construction, occur as a result of a breakdown of a form of (Galilean) relativity, associated with the underlying gauge symmetry. For this reason, in the absence of residual perturbations, effectively, it is impossible to perform a measurement that preferentially distinguishes between the behavior of particles in the bulk region in one reference frame, relative to the comparable behavior of particles in a second, moving reference frame that can be related to the first reference frame through a rigid, Galilean transformation. The resulting symmetry leads to a large degeneracy associated with elastic processes that can transfer momentum, instantly, from the boundaries of the bulk region to the center of mass of the bulk. In the limit in which the resulting motion (of the bulk) involves a transfer of momentum $\Delta p$ that has associated with it a DeBroglie wavelength $\lambda_d = \frac{h}{\Delta p}$ ($h$=Planck's constant) that is constrained by the condition that $n\lambda_d = a$ ($n$=integer, $a$=lattice spacing between adjacent unit cells), a resonant process can take place, in which, effectively, a small fraction $\frac{\Delta p}{N}$ of $\Delta p$ can be transferred instantly to the center-of-mass of each unit cell in the lattice. (Here, $N$ is the total number of unit cells in the lattice.) Historically, in infinitely-repeating lattices, an alternative picture has evolved. In it, the associated resonant processes (which will be referred to as Umklapp processes [Ashcroft & Mermin 1976B, Peierls 1929]) involve interactions in which momentum is not conserved between "quasi-particles". Instead, momentum is transferred to the lattice elastically. Although momentum conservation requires that quantitative bounds exist for the amounts of momentum that can be transferred from a crystal lattice to a surface or interface (and vice-versa) through these kinds processes, traditionally, in models in which the lattice is infinitely-repeating and periodic, these bounds have been poorly defined.

In fact, in finite solids, at low, but finite T, precise, size-dependent bounds can be identified. In particular, although in larger crystals, collisions with phonons tend to reduce the magnitudes of the associated effects, in smaller crystals (or in optical lattices), this is not the case. In the case of palladium-deuteride (PdD) and palladium-hydride (PdH), the effects can be quite large and can lead to coherent forms of interaction. In particular, when an electric field is applied for a sufficiently long period of time, in these systems, forms of insulating-conducting transitions can take place, in which the charge carriers are H- or D- ions. These processes are similar to the transitions suggested by Zener (1934) that could result when an electric field is applied to an insulator for a long time. As opposed to the situation in which "electrical conduction" through "Zener/Electrical breakdown" can occur (which is the name that is used for the effect that Zener suggested), because the charge carriers in PdH and PdD can be ions, an analogous process (which will be referred to in this paper as "ionic conduction" through "Zener/Ionic breakdown") can take place. In the case of PdD, since D-ions (deuterons) are bosons, the "ionic conduction" technically is a form "super-conduction," and the Zener/Ionic breakdown process refers to a form of insulator-superconductor transition. However, since the amount of ionic charge involved is small, the important effects, conventionally associated with superconductivity, are so small that they probably cannot be observed.

The range of time-scales (which varies between weeks and fractions of a second) for initiating this kind of effect appear to be consistent with the comparable range of incubation times that have been observed ( Arata & Zhang 1995 ; 1997B ; 1999A ; 2000A ; 2000B ; 2005 ) to be required before anomalous heat is initiated during the prolonged electrolysis of $D_2O$ by PdD. The fact that the shortest time-scales required for Zener/Ionic breakdown and the observation of Excess Heat both occur in PdD crystals that have characteristic dimensions ~10's of nm suggests that Excess Heat is being triggered by Zener/Ionic breakdown.

Including the Introduction, the paper contains five sections. In the next section, background material about broken gauge symmetry is presented. In the third section, rules about general properties of the ground state (GS) and lowest-lying excitations are presented (including a formal proof of a generalized form of Bloch's theorem). In the fourth section, a new, general framework (which will be referred to as Generalized Multiple Scattering Theory) is presented for identifying critical time- and length- scales associated with collisions, resulting from finite size effects, and implicit coherent effects. In this section, this formalism is applied in a particular limit (involving the onset of broken gauge symmetry) associated with coherent effects that become most pronounced either at finite, but vanishing-ly small T (in larger crystals), or in situations involving smaller crystals in which an external force acts on many particles at once (as in an Umklapp process) in a nearly perfectly rigid fashion, and bounds associated with the potential forms of coherent transfer of momentum (through Umklapp processes) from interior (bulk) regions of the lattice to regions outside the bulk are presented. Also, the section includes a formal, rigorous argument that quantifies limitations of Zener/Electronic breakdown and related phenomena, including, Bloch oscillations, in optical (and other) lattices, and the new effect (alluded to above) of Zener/Ionic breakdown in nano-scale PdH and PdD crystals. This section concludes with a concrete example that illustrates the usefulness of Generalized Multiple Scattering Theory, involving the identification of critical length, and time scales, associated with finite size, in the problem of measuring the gradient of the gravitational force. The final section provides a brief summary of the key results of the paper.

## II Background Concerning Conduction Near Absolute Zero

Although the semiclassical theory of electron dynamics and conduction (Ashcroft & Mermin 1976C) has been widely used, it has no formal justification, based on a microscopic theory. In fact, beginning from the GS of a many-body system in which the density of all of the particles, approximately periodically repeats itself, over some finite region of space, it is possible to formally justify when this treatment of conductivity applies for electrons, potentially for lighter ions (such as protons, deuterons and tritons), and/or, as discussed below, (in the case of optical lattices) neutral atoms, based on a microscopic theory, in a finite crystal. The essential physics is that when the GS is ordered, the lowest energy forms of interaction are initiated in the periodically ordered (bulk) regions but induce disorder through processes that are perfectly elastic, either within the initial region (or throughout most of it), while effectively transferring the momentum of the process instantly to the center-of-mass (CM) of the bulk, as a whole (as in the response of a solid to a gamma ray in the Mossbauer effect). At higher temperatures, alternative effects can occur, in which (ion and electron) charge, or (in the case of neutral atoms) a flux of atoms is transported to the boundaries of the lattice. (In this context, and henceforth, the term ion refers to the nucleus of hydrogen or one of its isotopes.)

These conclusions follow from two key features of an ordered solid, near T=0: 1.) The motion of bulk regions of the solid (which are regions in which charge is always conserved, and net changes in charge and total charge always vanish) or optical lattice (in which the number of atoms is conserved), relative to locations outside of the bulk region (in which charge, or atom number, is allowed to change), can never be determined without introducing some form of external perturbation; and 2.) The GS many-body wave function in bulk regions has minimal overlap with excited states that couple to outside forces and perturbations, involving non-bulk regions where charge imbalance (or, in the case of neutral atoms, in an optical lattice, a loss of particles) is allowed to take place. The first characteristic implies a form of symmetry: In a finite lattice, it is never possible to determine the constant zero of (kinetic) energy or momentum of the bulk region, relative to non-bulk regions, but, to determine the GS, in bulk regions, it can be assumed energy is conserved. Then, it is impossible to determine if bulk region particles are at rest or in motion.

This symmetry has important consequences: A.) Because rigid translations of the bulk uniformly shift the momentum of each bulk region particle, these kinds of translations do not alter the relative interactions between (or fluxes involving) different particles within the bulk region. B) The GS is defined in a preferred reference frame, in which the balance between outside forces defines the zeroes of energy and momentum. An important point is that the relationship between the velocity **v** and momentum **p**, of a charged particle, possessing charge e and mass m, is not **p**=m**v**. The precise relationship is m**v**=**p**-e/c**A**, where **A** (the vector potential) is defined by the magnetic field, **B** (through the relationship $\nabla \times \mathbf{A}=\mathbf{B}$) and Maxwell's Equations. In particular, quantum mechanically, **p** and **A** can both change instantaneously by the same amount, even discontinuously, at any location, without changing the value of v of any particular particle. The example of a Mossbauer-like, Galilean translation further illustrates subtleties associate with the effect. In this kind of situation no net accumulation of charge occurs. Then, it is impossible to determine if the bulk region is in motion or at rest, and, in the reference frame of an outside observer who moves with velocity -$V_{cm}$, the bulk region appears to move rigidly with total momentum $P_{cm}=MV_{cm}$ (M= mass of bulk solid), while in the frame in which both observer and bulk region are stationary, $P_{cm}=0$. Because the transformation is rigid, the relationship between the wave functions $\Psi_{bulk}(P_{cm}=MV_{cm})$ and $\Psi_{bulk}(P_{cm}=0)$ in the different frames involves a simple, change in phase:

$$\Psi_{bulk}(P_{cm}=MV_{cm}) = e^{\frac{iMV_{cm}\bullet R_{cm}}{\hbar}} \Psi_{bulk}(P_{cm}=0) = \prod_{i=1,N_{total}} e^{ik_i \bullet r_i} \Psi_{bulk}, (1)$$

where $R_{cm}$ is the position of the CM, $\hbar k_i = m_i V_{cm}$ is the momentum that a neutral particle with mass $m_i$ would have in the reference frame in which all particles move with velocity $V_{cm}$. The subtleties occur in this last example because near T=0 it is never possible to determine whether or not the "bulk " is in motion or at rest, or whether or not its particles really are "neutral", quantum mechanically. Eq. 1 has important consequences. In particular, non-local forms of coherence can occur, in which it is possible to maintain a T=0 environment, with no interaction, whatsoever in bulk regions, provided the entire bulk region moves all at once, but with different amounts of momentum (associated with different particles, as in Eq. 1).

# III Generalization of Bloch's Theorem in Finite Lattices, Near T=0

Depending on whether the bulk is in motion or at rest, relative to non-bulk regions, wherever a particular coordinate r associated with a particle of mass m, and charge e, appears in an expression involving the total many-body wave function $\Psi$, the expression should be multiplied by a pre-factor of the form, $\exp(i\mathbf{k} \bullet \mathbf{r})$, where $\hbar \mathbf{k} = mV_{cm} + \langle e/c\mathbf{A}(\mathbf{r})\rangle$, and $\langle e/c\mathbf{A}(\mathbf{r})\rangle = \hbar \mathbf{k_o} = \mathbf{p_o}$ is the average, minimum (zero) of the momentum of the particle. Here, degeneracy can occur because A (as well as the value of $k_o$) is never uniquely defined since the gradient of an arbitrary function can always be added to A, without altering the value of the magnetic field.

However, near T=0, it is also required that in the presence of a finite lattice, the associated forms of interaction between many-body states involve minimal, mutual overlap, in bulk regions. Thus, it follows that a discrete form of the degeneracy is involved in which, a priori, any one of the possible states, associated with a particular wave-vector, can couple to an alternative state associated with a different wave-vector, through an outside perturbation. As a consequence, through any of the possible symmetry operations (in which the bulk is translated rigidly), the value of $P_{cm}$ associated with one such translation can only differ from the comparable momentum of a second translation by the product of $\hbar$ and one of the wave-vectors within the First Brillouin Zone (defined by Born VonKarman boundary conditions [ Ashcroft & Mermin 1976D ] of the finite crystal). This result follows by considering the potential forms of interaction between the GS and possible, low-lying forms of excited states.

In particular, as a function of time t, for the lowest energy (GS) many-body wave function $\Psi_{GS}(r_1,....,r_n,t)$ to have minimal coupling with outside processes, its overlap with any other many-body state $\Psi'(r_1,....,r_n,t)$ must be minimized and remain constant. A requirement for this to occur is:

$$\frac{\partial \langle \Psi'|\Psi_{GS}\rangle}{\partial t} = \iiint d^3r_1...d^3r_n \frac{\partial(\Psi'^* \Psi_{GS})}{\partial t}$$

$$= -\int d^3r \nabla \bullet \langle \Psi'|v(r)|\Psi_{GS}\rangle + \langle \Psi'|\frac{V-V'}{i\hbar}|\Psi_{GS}\rangle = 0, \quad (2)$$

where terms in the second equality are defined by the many-body Schroedinger equations of $\Psi'$ and $\Psi_{GS}$. In general, the associated integrations are unrestricted. To minimize overlap in "bulk regions", unrestricted integrations over all of the coordinates in the multi-dimensional integral, term by term, can be restricted to regions in the bulk, based on the criteria that to find a possible GS, the associated overlap between this state and other states in the bulk region be minimized. In this limited context, by restricting states to have minimal overlap with $\Psi_{GS}$, additional restrictions are imposed on $\Psi_{GS}$ (including the possibility that *A* and/or $\nabla \Psi_{GS}$, *A'* and/or $\nabla \Psi'$, or all of these quantities change discontinuously, at the boundaries of the bulk region). Then, the associated analysis proceeds by restricting the multi-dimensional integrations in Eq. 2, exclusively to the bulk region. Also, in Eq. 2, $\langle \Psi'|v(r)|\Psi_{GS}\rangle$ is the matrix element associated with the contribution to the (many-body) particle velocity operator v, defined by its overlap with the states $\Psi'$ and $\Psi_{GS}$:

$$\langle \Psi'|v(r)|\Psi_{GS}\rangle = \sum_j \iiint d^3r_1..d^3r_n \delta^3(r-r_j) \frac{1}{m_j}(\frac{\hbar}{2i}[\Psi'^* \nabla_{r_j}\Psi_{GS} - \nabla_{r_j}\Psi'^* \Psi_{GS}] - \frac{e_j}{c}\Psi'^* A_{eff}(r_j)\psi_{GS}),$$

(3)

where $A_{eff}(r)=(A(r)+A'(r))/2$ is the arithmetic mean between the vector potential A'(r) associated with the state $\Psi'$ and the comparable vector potential A(r), associated with the state $\Psi_{GS}$, and the final term in Eq. 2 is defined by the difference between the many-body potential energy associated with states $\Psi'$ and $\Psi_{GS}$. In particular, this last term is given by

$$< \Psi' | \frac{V-V'}{i\hbar} | \Psi_{GS} > = < \Psi' | \frac{V_{em} - V'_{em}}{i\hbar} | \Psi_{GS} > + < \Psi' | \frac{V_s - V'_s}{i\hbar} | \Psi_{GS} >, \quad (4)$$

where $< \Psi' | V_{em} - V'_{em} | \Psi_{GS} >$ is the difference in electromagnetic potentials associated with coupling between the vector potentials A'(r) and A(r),

$$< \Psi' | V_{em} - V'_{em} | \Psi_{GS} > = \int d^3 r \frac{(A(r) - A'(r))}{c} \bullet J(r), \quad (5)$$

defined by the associated current J(r),

$$< \Psi' | J(r) | \Psi_{GS} > = \sum_j \iiint d^3 r_1 .. d^3 r_n \delta^3(r-r_j) \frac{e_j}{m_j} (\frac{\hbar}{2i} [\Psi'^* \nabla_{r_j} \Psi_{GS} - \nabla_{r_j} \Psi'^* \Psi_{GS}] - \frac{e_j}{c} \Psi'^* A_{eff}(r_j) \psi_{GS}),$$

and (in Eq. 4), the remaining contribution to the difference in potential energy is defined by any change in electrostatic or other (for example, inertial) contribution to the energy, associated with the transition from $\Psi'$ (where the non-electrodynamic portion of the potential energy is $V_s$') to $\Psi_{GS}$ (which has a corresponding non-electrodynamic potential energy $V_s$).

    Eq. 2 vanishes identically whenever the energies associated with $\Psi_{GS}$ and $\Psi'$ are the same. When $\Psi_{GS}$ has minimal coupling to the bulk, Eq. 2 holds identically, outside the bulk, provided all of the external forces vanish and the total internal flux of all particles into and away from the bulk region also vanishes. Thus, if the flux of particles, across all boundaries in the bulk vanishes, and the energies of the different states are the same within the bulk region, it follows from Eq. 2 that,

$$\int d^3 r \nabla \bullet < \Psi' | v(r) | \Psi_{GS} > = \int_{\partial V} dS \, \hat{n} \bullet < \Psi' | v(r) | \Psi_{GS} > = \frac{i}{\hbar} \iiint_V d^3 r_1 ... d^3 r_n \Psi'^* (V-V') \Psi_{GS} = 0 \quad (6)$$

where the integration in the final term extends over the bulk region, and the surface integral (associated with v(r) ) extends over the boundary of the bulk region. In principle, although this surface integral includes separate contributions from regions where v may become discontinuous (which are allowed to occur whenever V-V' becomes singular), for the purpose of identifying the GS, $\Psi_{GS}$ and $\Psi'$ can be selected in such a way that V-V' is never singular. Then, a necessary and sufficient condition to guarantee that the left-side (LS) of Eq. 2 vanishes within some volume, defined by a set of boundary planes, in which each point r on one boundary plane is related to a point r' on a second boundary plane, by one of three (linearly independent) vectors, $\vec{L}_\alpha$, is that

$$v(r + \vec{L}_\alpha) = v(r') = v(r). \quad (7)$$

In the limit in which $\Psi_{GS}$ and $\Psi'$ are identically the same in the bulk region (but are allowed to be different outside the bulk), and $A_{eff}$ equals a constant, Eq. 7 holds if and only if when for each coordinate $r_i$ that is evaluated on a boundary at $r' = r_i + \vec{L}_\alpha$, then

$$| \Psi_{GS}(r_1,..,r_i + \vec{L}_\alpha..,r_n) |^2 = | \Psi_{GS}(r_1,..,r_i..,r_n) |^2, \quad (8)$$

and

$$\frac{\partial \ln \Psi_{GS}(r_1,..,r_i + \vec{L}_\alpha..,r_n)}{\partial r_i^m} = \frac{\partial \ln \Psi_{GS}(r_1,..,r_i..,r_n)}{\partial r_i^m}, \quad (9)$$

where $r_i^m$ is the x, y or z component (for m=1,2, or 3) of the coordinate $\mathbf{r_i}$. The general solution of Eqs. 8 and 9 is

$$\Psi_{GS}(r_1,..,r_i + \vec{L}_\alpha..,r_n) = \lambda_i \Psi_{GS}(r_1,..,r_i..,r_n), \; | \lambda_i | = 1. \quad (10)$$

Eq. 10 is a generalization of Bloch's theorem, for finite lattices that holds whenever it is possible to define boundaries through the three displacement vectors, $\vec{L}_\alpha$, for a GS that obeys Eq. 6. In particular, when Eq. 10 holds over distances that are smaller, it also holds when smaller primitive vectors $\vec{b}_\alpha$ are used. When $\vec{b}_\alpha = \frac{\vec{L}_\alpha}{2N_\alpha}$ (where $2N_\alpha$ = number of unit cells between boundaries, defined by Eq. 9), it follows from Eq. 10 that

$$\lambda_i(\vec{L}_\alpha) = \lambda_i(2N_\alpha \vec{b}_\alpha) = \lambda_i(\vec{b}_\alpha)^{2N_\alpha}. \tag{11}$$

Since the right-side (RS) of Eq. 9 is independent of the initial displacement that appears on the leftside (LS), the LS must be stable with respect to additional variations in either $\vec{b}_\alpha$ or $\vec{L}_\alpha$ (either through infinitesimal variations in the scale of either vector, or through a rigid translation of the lattice). In general, the rightside (RS) of Eq. 5 does not vanish. But a rigid translation of the lattice occurs when $\Psi'$ satisfies an alternative version of Eq. 10 associated with an alternative reference frame, in which $\lambda_i$ is replaced with a different eigenvalue $\lambda_i'$ and the vector potential A is replaced with a different vector potential A' (where A and A' differ by a uniform constant, throughout the bulk). Because the associated transformation can be applied through an infinitesimally small displacement, it follows that $\nabla_{\vec{b}_\alpha} \lambda(R_n)$ is independent of $\vec{b}_\alpha$. But, in order to have minimal overlap with the GS, the allowable states associated with alternative values A' of the vector potential must be selected so that the RS and LS of Eq. 5 vanish. This requirement leads to the constraint that A' be selected, relative to A, in such a way that

$$\ln(\frac{\lambda_i(b_\alpha)}{\lambda_i'(b_\alpha)}) = i\frac{\pi n}{N_\alpha} = ik \bullet b_\alpha, \tag{12a}$$

where n is an integer, and k is one of the discrete vectors defined by the finite lattice. Here, for the different states (associated with the different values of $\lambda_i(b_\alpha)$ and $\lambda_i'(b_\alpha)$ in Eq. 12a) to be orthogonal (which is necessary for minimal overlap between $\Psi'$ and $\Psi_{GS}$), each possible value of $k \equiv \vec{k}_i$ on the far RS of the equation is required to be in the First Brillouin Zone, defined by

$$\vec{k}_i = \sum_{\alpha=1,3} \frac{i_\alpha \vec{g}_\alpha}{2N_\alpha} \quad ;-N_\alpha \leq i_\alpha \leq N_\alpha - 1, \ N = 8 N_1 N_2 N_3, \tag{12b}$$

where the three reciprocal lattice primitive vectors $\vec{g}_\alpha$ are constructed using,

$$\vec{g}_\alpha \bullet \vec{b}_{\alpha'} = 2\pi \delta_\alpha^{\alpha'}, \tag{12c}$$

where $\delta_\alpha^{\alpha'} = 1$ when $\alpha = \alpha'$, $\delta_\alpha^{\alpha'} = 0$ when $\alpha \neq \alpha'$. Eqs.12a-c have additional implications: Because $\lambda_i(b_\alpha)$ and $\lambda_i'(b_\alpha)$ in Eq. 12a are different but are related to each other through a uniform shift in the vector potential, associated with the requirement that the total flux, in Eq. 5 vanish, effectively, the equation establishes a preferential reference frame associated with the bulk GS and low-lying excited states (defined through Eqs. 10-11). In particular, in this frame, the zero of momentum of each state is selected so that the overlap between any two states possessing different wave-vectors vanishes (as a consequence of being orthogonal to each other). This requirement leads to the result that when each state is selected to be initially in the First Brillouin zone (which is appropriate when the GS is at rest, and the remaining states are related to each other and the GS through rigid, Galilean transformations), the difference in wave-vectors (as a result of preferentially picking the vector potential) always is required to be in the First Brillouin zone. The associated construction fixes the relative difference between the zeroes of momentum of the GS and the low-lying excited states, and zero of momentum of one low-lying excited state, relative to a second low-lying excited state.

The associated restriction on the relative differences in zeroes of momentum also restricts the implicit degeneracy associated with Umklapp processes in infinitely-repeating, periodic lattices. In particular, in infinitely-repeating periodic lattices, Born VonKarman boundary conditions are imposed, based on arbitrary (but large) values of $|\bar{L}_\alpha|$. But since these values are arbitrary, it is assumed that larger values can be employed. The arbitrariness in length-scale would imply that an integer multiple of $2N_\alpha$ could be arbitrarily added to n on the RS of the first equality of Eq. 12a. However, in a finite lattice, the displacement vectors, $\bar{L}_\alpha$, are determined unambiguously by the requirement of vanishing GS flux across a well-defined boundary (as in Eq. 7) and that states that are related to the GS by rigid translations of the bulk be orthogonal to the GS. In particular, when each $\vec{k}_i$ is in the First Brillouin Zone, many-body states associated with different sets of wave-vectors become orthogonal, provided the finite (discrete) Fourier transform, that is used to define the Bravais Lattice vectors $R_n$, involves the same, discrete set of integers (so that each of the possible integer values $n_\alpha$ that are used in forming the lattice vectors $R_n = n_1 \vec{b}_1 + n_2 \vec{b}_2 + n_3 \vec{b}_3$ are constrained by the inequality, $-N_\alpha \le n_\alpha \le N_\alpha - 1$ ).

Since $|\lambda_i| = |\lambda_i'| = 1$ in Eq. 12, it follows that both $\Psi_{GS}$ and $\Psi'$ can be written using a common functional form $\Psi$, in which the dependence on changes in the phase associated with either eigenvalue (as in Eq. 12) occur through a plane-wave that changes as any of the coordinates is displaced by a Bravais vector $R_n$, and through a second function u that is periodic with respect to translations of any of its coordinates by $R_n$; i.e. for $\Psi = \Psi'$ or $\Psi = \Psi_{GS}$, $\Psi$ can be written as

$$\Psi(r_1,....,r_n) = e^{i\sum_j k_j \bullet r_j} u(r_1,.....,r_n), \quad (13)$$

where $u(r_1,.....,r_n) = u(r_1,..r_i + R_n...,r_n)$ for all coordinates $\mathbf{r_i}$. Because $\Psi_{GS}$ or $\Psi'$ can be written using Eq. 13, the gradient of $k_i \bullet r_i$ in the exponential factor simply alters the value of A or A' through a (trivial) gauge transformation (associated with changing the value of $\mathbf{p_o}$ ), for each particle in the many-body Schroedinger equation of the one state (associated with $V_s = V_s'$ ) that differs from the other only through a change in $\mathbf{p_o}$ .

Thus, a large degree of symmetry exists, in which one, two,…n, Bravais translations, in principle could be performed, in which the energy is not changed, while the value of $\mathbf{p_o}$ associated with the coordinate of one (or more) particles is shifted relative to another. (This is the origin of the generalized, double Bloch symmetry that has been previously used [Chubb, T.A. 2005, Chubb & Chubb 2001].) In fact, outside forces constrain the lattice and break the associated degeneracy. As alluded to above, the lowest energy processes involve situations in which, in the bulk region, the state describes a configuration of particles that is neutral (on the average, in each unit cell), and in which all particles move with a common, velocity $V_{cm}$. In this limit, $\sum_i k_i \bullet r_i = \frac{V_{cm}}{\hbar} \sum_i m_i \bullet r_i = \frac{MV_{cm} \bullet R_{cm}}{\hbar}$, and the vector potential for each charged particle (measured relative to its constant zero ) either is unchanged, or all vector potentials are shifted by a common, uniform constant amount.

In practice, determining $\mathbf{p_o}$ for the GS (and low-lying excited states) for each kind of particle (in a charged many-body system) is formidable on a microscopic scale in finite crystals because as charge begins to accumulate, potentially large variations in electric field and charge density can occur. In Optical Lattices [Denschlag et al 2002, Deutsch & Jessen 1998], the comparable problem is not as severe because the lattices are created artificially, and the limiting forms of disorder can be externally controlled and monitored. In either case, in general, it is impossible to identify GS properties based on the kind of (simpler) rules that apply in bulk crystals, and calculations that include specific information about non-bulk regions (near surfaces or interfaces, in solids, and at the boundaries of the lattices in other cases) are required. In larger lattices, in solids, asymptotically, it is possible to understand, at least in an average sense, how by averaging many terms, a number of important approximate aspects of the associated coupling can occur. The resulting perturbations break the degeneracy of the lowest energy states in the "bulk solid" by establishing a preferential reference frame that fixes the value of $p_o$, relative to each of the allowable values of the momentum $p_j = \hbar k_j - P_{cm} + p_o$, associated with different reference frames, defined by the possible rigid Galilean transformations, as in Eq. 1. Then for the purpose of labeling eigenstates, each coordinate in each many-body wave function can be assigned one of the possible values of $p_j = \hbar k_j$, by requiring that $p_o = P_{cm} = \sum_j p_j$. This also fixes the zero of energy $\varepsilon_j \equiv \varepsilon(p_j)$, associated with the particular reference frame, as a function of $p_o$; i.e. $\varepsilon_j = \varepsilon(\Delta p_j + p_o), (\Delta p_j = p_j - p_o)$.

Once the zero of momentum, $p_o$, of the solid (as a whole) is fixed, all states, associated with symmetries that are allowed in Eq. 10 can become possible. In particular, in the most general situation, it is not possible to tell if the bulk is in motion or at rest. Then, by convention, $p_o = 0$, and $p_j = \hbar k_j = \Delta p_j$ is the minimum value of momentum associated with the dependence of $\Psi_{GS}$ on the coordinate $r_j$, that asymptotically can be related to a particular particle or (in situations involving correlation) a collection of particles. Each value of $p_j$, in turn, is fixed by the average variation in $A(r_j)$ that can result from its overlap with $\Psi_{GS}$ (at the location $r_j$). In general, every coordinate in $\Psi_{GS}$ should be treated as having a separate zero of kinetic energy $V_j$. In practice, each value of $V_j$ is defined (for example, using Eq. 3) through implicit matching conditions that require that changes in the zero of energy associated with the classical turning point for a particular coordinate $r_j$ (defined by $\frac{1}{2}mv(r_j)^2 = 0$) *can* result in discontinuities in the logarithmic derivatives of $\Psi_{GS}$ and/or $\Psi'$. Then, as a consequence of momentum conservation, at the boundaries of the bulk, the problem of determining $V_j$ in the bulk is equivalent to the problem of solving an equivalent minimization problem (which can be expressed in terms of a well-defined Rayleigh-Ritz variational procedure) for each eigenvalue $\varepsilon_j$ (which can be defined as a generalized form of band state energy) associated with the many-body Schroedinger equation. In particular, to determine the GS, it is possible (and consistent with energy minimization) to define each value of $V_j$, using $V_j = \varepsilon_j$, and to require that $E_{GS} = \sum_j \varepsilon_j$.

The solution of the many-body Schroedinger Equation for each value of $\varepsilon_j$ requires detailed information about particle flux at the boundaries of the bulk and additional information (when correlation is present) associated with particle exchange symmetry, including effects involving the possible exchange of internal quantum numbers (such as magnetic spin). But it is possible to use approximate boundary conditions in order to understand the underlying physics associated with dynamical transport in finite lattices. In particular, considerable progress can be made by requiring that the functional form of the most general many-body wave function that is used to describe the GS (or the lowest-lying excited states) asymptotically approach the functional form that applies in the independent particle limit in non-bulk regions either far from the bulk or (provided the wave function asymptotically approaches the associated functional form far from the boundaries and is non-vanishing and continuously differentiable between the boundary and locations far from the boundary), at the boundaries of the bulk. Then, asymptotically, in these regions, the functional forms associated with $\Psi_{GS}$ and the low-lying excited states of the bulk can be expressed using anti-symmetrized (symmetrized) sums of products of single particle fermion (boson) wave functions, and it is possible to identify and order the associated eigenvalues $\varepsilon_j$, based on the conventional nomenclature (associated with "occupied" and "unoccupied" states) of the independent, single particle picture. As a consequence, values of $\varepsilon_j$ involving "occupied states" can be distinguished from "unoccupied states" (in $\Psi_{GS}$ and $\Psi'$). Similarly, hole-like states can be distinguished from particle-like states ( Chubb 2005A ; 2005D ).

By requiring that $\langle \Psi_{GS} | \Psi_{GS} \rangle$ be constant and stable with respect to infinitesimal variations in the value of $p_o$, it is possible to derive a generalization of the semiclassical dynamics and transport theory of charged (band state) particles (electrons and ions). Here, provided externally applied forces F and charge vary sufficiently slowly in the external regions, the gradient of each value of $\varepsilon(p_j)$ with respect to $p_j$, for a particular particle, identically equals the expectation value of the contribution from coordinate $r_j$ to the velocity operator v, averaged over the bulk and surface regions. Also, provided F varies sufficiently slowly, the associated changes ( $\Delta p_j$ and $\Delta p_o$ ) in $p_j$ and $p_o$ obey $\Delta p_j = \Delta p_o = \int F \, dt$. Thus, the changes in wave-vector $\Delta k = \frac{\Delta p}{\hbar}$, as functions of time, conventionally associated with each band energy in the semiclassical theory ( Ashcroft & Mermin 1976C), can be rigorously interpreted as shifts in both the momentum and zero of momentum of each state. Details about this and its implications on transport phenomena are discussed elsewhere ( Chubb 2005A ; 2005D ). An important point is that the lowest-lying excitations involve coupling between different generalized Bloch states (as in Eq. 13) with outside perturbations. In particular, because of this fact, with increasing crystal size, it might seem plausible that the dominant forms of coupling at low but finite T involve resonant (Umklapp) processes that approximately conserve energy and momentum in the bulk. In the next section, an argument that involves a generalization of multiple scattering theory (MST) is presented and used to establish quantitative bounds associated with critical length and time scales for justifying this intuitively appealing idea. Also in this section, a comparable application of this generalization of MST associated with neutral atoms is presented, and a number of additional important effects (associated with coherence and the onset of broken gauge symmetry) are presented that have bearing on an important practical problem, precision measurements of gravity, using ultra cold atoms, and possibly on a second important effect: the evolution of anomalous forms of heat that have been observed during the prolonged electrolytic loading of deuterium (D) into PdD lattices.

# IV Onset of Broken Gauge Symmetry and Coherent Effects in Finite Lattices at Vanishingly Small T

UmKlapp-like processes are the dominant low-energy fluctuations because they preserve translation symmetry in the bulk. This fact can be justified formally, using the Lippman-Schwinger equation(LSE) ( Levine 1999 , Schwinger 1990 ), or, equivalently, from R-Matrix Theory (Burke & Berrington 1993). In particular, in general, if $\Psi_o$ is the wave function that describes any of the initial, possible, low-lying, many-body states that are associated with either a particular region of space, or some finite portion of the energy (and momentum) spectrum, or both, through excitation and/or de-excitation of the initial state, an outside perturbation $\Delta V$ will induce overlap between $\Psi_o$ and one or more states associated with a different ("forbidden") region of space or portion of the spectrum. It follows from the LSE (as well as R-Matrix theory) that the general rate of reaction $R = \lim_{t \to \infty} \dfrac{\partial \int d^{3n}r \left| \Psi_o^+(r_1,...,r_n,t) - \Psi_o^-(r_1,...,r_n,t) \right|^2}{\partial t}$ (or, equivalently, lifetime $\tau \equiv \dfrac{1}{R}$ ), defined by the long time limit of the difference between the asymptotic form that $\Psi_o$ approaches in the distant future ($\Psi_o^+$) and past ($\Psi_o^-$) that results from excitation/de-excitation of the system, is given by (Schwinger 1990):

$$R \equiv \frac{1}{\tau} = \frac{2\pi}{\hbar} < \Psi_o | \Delta V \delta(E-H) \Delta V | \Psi_o >$$
$$\equiv \frac{2\pi}{\hbar} \sum_F \delta(E - E_{exact}(C_F)) |<\Psi_o | \Delta V | \Psi_{exact}(C_F)>|^2, \qquad (14)$$

where $\Psi_{exact}(C_F)$ is any many-body state, possessing a particular configuration (denoted by $C_F$ ) of particles that solves the Schroedinger equation of the exact Hamiltonian, H, associated with the exact potential $V$, and $E_{exact}(C_F)$ is its energy.

When the perturbation preserves translation symmetry in the bulk region, $\Delta V$ vanishes except in non-bulk regions. For this case, the set of initial many-body states corresponds to the set of (degenerate) states associated with the bulk region, in the absence of overlap with non-bulk states, and the perturbation $\Delta V = V - V'$ induces the associated broken gauge symmetry that results from the overlap between these sets of states. But because of the energy-conserving delta function, in Eq. 14, in the associated summation over configurations, $\Psi_{exact}(C_F)$ and $\Psi_o$ both have the same energy E. But because, formally, the inner product between states that have the same energy is independent of time, Eq. 14 can be re-written, using Eqs. 1-3. In particular, Eqs 1-3 imply that

$$i\hbar \int d^3 r \nabla \bullet <\Psi_o | v(r) | \Psi_{exact}(C_F)> = <\Psi_o | \Delta V | \Psi_{exact}(C_F)>$$
$$= i\hbar \sum_\alpha \int_{S_\alpha} d^2 r_\alpha <\Psi_o | v(r_\alpha) | \Psi_{exact}(C_F)>. \qquad (15)$$

Here, we have used Gauss's law and the definition of $v(r)$ ( given in Eq. 3) to convert the volume integral on the LS of the equation (in a formal sense) into a sum of surface integrals. The sum extends over the boundaries of all of the regions where the divergence on the LS, effectively, can become discontinuous. In particular, Eq. 14, in principle, can be applied to situations, in which, locally, violations in particle number and/or momentum conservation in the initial state are allowed to take place. This can occur as a result of discontinuities in momentum, vector potential or both, that can result from any broken gauge symmetry, over a region of any size. The summation on the RS extends over the boundaries of all of these "forbidden" regions.

In principle, each term in this summation can be viewed as an integral representation of a contribution to the matrix element of a particular scattering event. Substituting Eq. 15 into Eq. 14, we can relate the total reaction rate R to the square of a matrix element, derived exclusively from the total net flux of particles into and away from the "allowed" region (associated with any of the allowable states that have particular energies or momentum and/or are confined to a particular region):

$$R \equiv \frac{1}{\tau} = 2\pi\hbar \sum_F \delta(E - E_{exact}(C_F)) | \sum_\alpha \int_{S_\alpha} d^2 r_\alpha < \Psi_o | v(r_\alpha) | \Psi_{exact}(C_F) > |^2, \quad (16)$$

In its most general form, the forbidden region boundaries can refer to locations where any "collision" (associated with the regions involving discontinuous changes in momentum) can occur.

Eq. 16 is a many-body version of a general set of equations that have been used either to describe scattering or to compute bound state eigenfunctions and eigenvalues of low energy electrons in solids. In particular, when the equation involves only a single particle, it reduces to the familiar rate expression that is used in multiple scattering theory ( Gonis & Butler 2000 ). The generalization, given by Eq. 16, which includes many particles, has been referred to as Generalized Multiple Scattering Theory ( Chubb & Chubb 2000 ). Eq. 16, in principle, can be used to identify bounds for particular processes, provided it is possible to make plausible assumptions about the many-body wave function, its gradient, and the behavior of the vector potential, in various regions of space. In situations involving reduced values of R, associated with near ground state configurations, it is possible to use Eq. 16 to identify the impact of boundaries on potential forms of coherence (especially broken gauge symmetry). Beginning from Eq. 16, it is also possible to identify the onset of new forms of broken gauge symmetry, associated with possible ionic conduction that can occur in situations involving high-loading (defined by the limit in which x→1 in PdH$_x$ ) of atomic hydrogen (H) and/or its isotopes into palladium (Pd), in finite size crystals ( Chubb & Chubb 2000 ). The associated forms of coherent (resonant) interactions occur because of the large degeneracies that result from the energy conserving delta function when the energy E is small and/or when the dominant scattering processes involve a small number of coherent interactions.

At finite, but low T, excitations associated with these degeneracies provide the dominant forms of interaction. As a consequence, in principle, beginning from the GS, it is possible to describe the time evolution of the lowest energy fluctuations of the system, by generalizing an earlier procedure ( Lipavsky et al 1986). Specifically, formally, based on the identity,

$2\pi\delta(x) = i \lim_{\varepsilon\to 0}[\frac{1}{x+i\varepsilon} - \frac{1}{x-i\varepsilon}]$, it is always possible to re-write Eq. 16, in the form,

$$R \equiv \frac{1}{\tau} = \hbar \lim_{\varepsilon\to 0} \sum_{\alpha,\alpha'} \int_{S_{\alpha'}} d^2 r_{\alpha}' \int_{S_\alpha} d^2 r_\alpha < \Psi_o | v(r_\alpha)(i) P_{\Psi(C_F)} [\frac{1}{E - H_{exact} + i\varepsilon} - \frac{1}{E - H_{exact} - i\varepsilon}] P_{\Psi(C_F)} v(r_\alpha') | \Psi_o >, \quad (17)$$

where, formally,

$$P_{\Psi(C_F)} = P_{\Psi(C_F)} P_{\Psi(C_F)} = \sum_F |\Psi_{exact}(C_F)\rangle\langle\Psi_{exact}(C_F)|, \quad (18)$$

represents the projection of the exact wave function onto the possible configurations that have appreciable overlap with the lowest-lying excited states, and $\frac{1}{E - H_{exact} + i\varepsilon}$ and $\frac{1}{E - H_{exact} - i\varepsilon}$, respectively, are potential representations of the inverse of the (many-body) operators defined by the differences ($E - H_{exact} + i\varepsilon$ and $E - H_{exact} - i\varepsilon$) between the complex energies $E + i\varepsilon$ and $E - i\varepsilon$ and the exact, many-body Hamiltonian $H_{exact}$ (in first quantized form) associated with the many-body Schroedinger equation. In the limit of finite, but vanishing-ly small T, beginning with large (but finite) values of τ, as a result of gauge symmetry (and the onset of broken gauge symmetry), Eq. 17 also can be used to identify and isolate a hierarchy of coherent Umklapp-like processes, associated with rigid, Galilean transformations, in finite lattices. In particular, to understand the origin and significance of these kinds of processes, general features of the associated forms of overlap and coupling (through broken gauge symmetry) to non-bulk effects can be identified, without averaging Eq. 17 over initial state energies (as opposed to situations, associated with higher values of T, where this kind of averaging is required).

This is possible because broken gauge symmetry is initiated through processes that reduce the degeneracy of the system, and the GS is required to have the lowest degeneracy and minimal overlap with other states that are related to it (in the absence of broken gauge symmetry) by rigid Galilean translations. Formally, in order to include the most general forms of processes that break gauge symmetry, all possible values of $k_i$, that, in the absence of broken gauge symmetry, are required to be degenerate, must be included. For this reason, the many-body wave functions for the lowest-lying excited states and the GS are all required to include effects associated with the variation in the zeroes of all possible CM momenta, for all of the wave-vectors in the First Brillouin zone, in each filled band, and for all of the states in the highest (lowest) occupied fermion (boson) band. This can be accomplished by allowing the summation ($i \sum_j k_j \bullet r_j$) in the exponential factor in Eq. 13 to include a sufficiently large, total number $N_T$ of coordinates to allow for variations in the zeroes of momentum of all of the particles in the bulk and with fluxes that can remove or introduce particles into the bulk (through matrix elements involving finite overlap with the GS in the bulk region) that can take place that preserve the degeneracy associated with rigid, translational symmetry. In particular, to describe the lowest-lying excitations, it is necessary to relate the individual coordinates ($r_1,....,r_{N_T}$) of the various particles in the many-body wave function to the coordinate $R_{cm}$ associated with the CM, the separation variables $r_{ij}$ that describe the difference in the position of each particle from the remaining particles, and the total mass M of the collection of particles. The general relationship between the general coordinates ($r_1,....,r_{N_T}$) and $R_{cm}$ and $r_{ij}$ is:

$$r_{ij} = r_i - r_j = -r_{ji}; \quad R_{cm} = \frac{\sum_{i=1,N_T} m_i r_i}{M} \quad . \quad (19)$$

It follows from Eq. 19 that

$$r_i = \sum_{j \neq i} \frac{m_j r_{ij}}{M} + R_{cm}; \quad r_j = r_i - r_{ij} \quad (j \neq i), \quad (20)$$

and that

$$\nabla_{r_i} - \frac{m_i}{M} \nabla_{R_{cm}} = \sum_{j \neq i} \nabla_{r_{ij}} = \nabla_{r_i - R_{cm}}; \quad \nabla_{r_j} - \frac{m_j}{M} \nabla_{R_{cm}} = -\nabla_{r_{ij}}, \quad j \neq i, \quad (21)$$

where the summations in Eqs. 19-21, in principle, should be carried out over all $N_T$ particles in the many-body system. In fact, the only contributions that are relevant in evaluating Eq. 17 occur from particles that have non-vanishing flux at the boundaries of the bulk region (and in what follows, $N_T$ will refer to the smaller number of particles that satisfy this constraint).

When the associated processes involve rigid translations (as in the potential overlap between the GS and/or the lowest energy excitations, associated with Umklapp-like processes), in the bulk region, the separation variables ($r_{ij}$) are held fixed. As a consequence, in the evaluation of the surface integrals, at the boundaries of the bulk,

$$\nabla_{r_i - R_{cm}} \Psi_{exact}(C_F)|_{boundary} = \nabla_{r_{ij}} \Psi_{exact}(C_F)|_{boundary} = \nabla_{r_i - R_{cm}} \Psi_o|_{boundary} = \nabla_{r_{ij}} \Psi_o|_{boundary} = 0, \quad (22a)$$

for each internal coordinate $r_i - R_{cm}$, associated with the position $r_i$ of a particle, relative to the CM coordinate $R_{cm}$ and each separation variable $r_{ij}$. It also follows from Eq. 21 that

$$\nabla_{r_i} \Psi_{exact}(C_F)|_{boundary} = \frac{m_i}{M} \nabla_{R_{cm}} \Psi_{exact}(C_F)|_{boundary}; \quad \nabla_{r_i} \Psi_o|_{boundary} = \frac{m_i}{M} \nabla_{R_{cm}} \Psi_o|_{boundary} \quad (22b)$$

Umklapp-like processes are dominant for sufficiently large lattices because: 1. In the absence of interaction (through fluxes of particles at the boundaries of the solid), it is never possible to determine if the bulk region is in motion or at rest; 2. Umklapp-like processes are equivalent to gauge transformations that preserve this symmetry, in the bulk. In particular, Eq. 1 is not valid in general (as a consequence of broken gauge symmetry). But it does hold (effectively) in the evaluation of contributions to the RS of Eq. 17, when the differences in velocity between each kind of indistinguishable particle associated with an initial and final state (GS or a low-lying excited state) preserve periodic order. This follows from the fact that by construction, when $\Psi_{exact}(C_F)$ and $\Psi_o$ are low-lying excited states, both states obey Eq. 10 and, as a consequence, are required to have a functional form, defined by Eq. 13. But because $u(r_1,.....,r_n) = u(r_1,..r_i + R_n...,r_n)$ in Eq. 13, it follows that

$$\frac{\partial u(r_1,.....,r_n)}{\partial R_{cm}} = 0. \quad (23)$$

It also follows from from Eqs. 19 and 20, that the phase factor in Eq. 13 can be re-expressed, using the expression

$$\sum_i k_i \bullet r_i = \sum_{i=1,N_T-1} \sum_{j=i+1,N_T} \frac{k_{ij} \bullet r_{ij}}{M} + k_{cm} \bullet R_{cm}, \quad (24)$$

where

$$k_{ij} = \frac{m_j k_i - m_i k_j}{M} \quad (25)$$

and

$$k_{cm} = \sum_{i=1,N_T} k_i. \quad (26)$$

Here, as in Eq. 13, each value of $k_i$ is in the First Brillouin Zone. Eqs. 23 and 24 imply that any wave function $\Psi(r_1,.....,r_n)$ that has a functional form given by Eq, 13 obeys the equation,

$$\frac{\hbar}{i} \frac{\partial \Psi(r_1,.....,r_n)}{\partial R_{cm}} = \hbar k_{cm} \Psi(r_1,.....,r_n). \quad (27a)$$

But when the transition involves a rigid translation of the bulk region, the only contributions that appear on the RS of Eq. 17, involve fluxes associated with $\nabla_{R_{cm}} \Psi_{exact}(C_F)$ and $\nabla_{R_{cm}} \Psi_o$. Then, because $\Psi_{exact}(C_F)$ and $\Psi_o$ satisfy Eq. 13, Eq. 27a holds when $\Psi(r_1,.....,r_n) = \Psi_{exact}(C_F)$ or $\Psi(r_1,.....,r_n) = \Psi_o$. As consequence, the contribution to the flux that results from a rigid translation in Eq. 17 from each indistinguishable particle can be derived using the transformation associated with Eq. 1.

By construction, as a consequence, the condition that no net flux be present allows us to always define the zero of momentum, associated with a particular value of the wave-vector, along a well-defined boundary, in such a way that different wave-vector states (as in Eq.12) be orthogonal. In a situation involving a rigid translation, however, a net flux can be present because only CM motion takes place, as a consequence of Eqs.22a-23, and it is no-longer possible to preferentially select the relative zeroes of momentum (through particular gauge transformations) in this manner. But relative to the initial bulk region, in all sub-lattices, except those that include the boundary, it is still possible to require that different wave-vector states (defined by the different sub-lattices) remain orthogonal (since in the integration over the boundary of any of these sub-lattices, there is no net particle flux). Also, no net flux of particles enters or leaves these regions since the lowest-lying excitations obey Bloch's theorem.

But in the regions that do include the boundary, states that have different sets of wave-vectors are no longer required to be orthogonal. Also, since charge can accumulate in these regions, the precise location of the physical boundary is not defined. In general, the many-body wave function must remain continuous and single-valued, but the momentum (associated with any of the coordinates) can become discontinuous, without altering the particle flux, on both sides of each boundary. Asymptotically, the lowest energy excitations occur when the longest wave-length perturbations induce variations in a small number of separation variables $r_{ij}$. This occurs when effectively, after the initial bulk region is translated rigidly by a small integer multiple of primitive vectors $\vec{b}_\alpha$, within a larger lattice, a small variation in one or more of the separation variables $r_{ij}$ is allowed to take place. In particular, initially, at all points within the bulk, for each of the coordinates, $(r_i, r_j, ...)$, $\nabla_{r_{ij}} \Psi_o = 0 = \nabla_{r_{ij}} \Psi(C_F) = \sum_{i \ne j} \nabla_{r_{ij}} \Psi_o = \nabla_{r_i - R_{cm}} \Psi_o$

$= \sum_{i \ne j} \nabla_{r_{ij}} \Psi(C_F) = \nabla_{r_i - R_{cm}} \Psi(C_F)$; while at the boundary, one or more values of $\nabla_{r_{ij}} \Psi(C_F)$ can become finite. The lowest energy configuration occurs when $\Psi(C_F) = \Psi_o$ everywhere in the bulk; while $\nabla_{r_{ij}} \Psi(C_F)$ is finite for one value of $r_{ij}$ at the boundary (which can only occur through a collision that leads to a discontinuity in $\nabla_{r_{ij}} \Psi(C_F)$). In the limit of finite $N$, it can be shown ( Chubb 2005D ) that minimal velocity flux change can occur from compensating, discontinuous gauge transformations, through discretely quantized changes $\Delta A$ in vector potential:

$$\oint \frac{e_j}{c} \Delta A \bullet dl = 2\pi m, \quad n = \text{integer}, \tag{27b}$$

where the integration is along any closed loop that encloses the boundary, and n, in principle, is an arbitrary integer. It also follows that, in this limit, $\Delta A$ is a constant, and that

$$N_\delta \frac{e_j}{c} \Delta A \bullet \vec{b}_\delta = \pi n_\delta, \, n_\delta = \text{integer}, \delta = 1,2,3; \delta \ne \alpha, \tag{28}$$

where $\vec{b}_\delta (\ne \vec{b}_\alpha)$ is one of the two (remaining) primitive vectors that can be used to define each of the lattice sites on the boundary of the lattice, and $N_\delta$ is defined by Eq. 12. As a consequence of Eq. 12b, it also follows that the lowest lying excitations are integer multiples of $\frac{e_j}{c} \Delta A = \frac{\hbar \vec{g}_\delta}{2N_\delta}$ and that they introduce small deviations in the average CM momentum, as discussed elsewhere ( Chubb 2005D ). By construction, the particle flux in Eq. 27b provides the smallest contribution to the total rate (in Eq. 17). The associated rigid, Galilean transformation, technically, is an Umklapp process, but for an elongated lattice, in which the initial primitive vector $\vec{b}_\delta$ is replaced by the vector $\vec{L}_\delta = 2N_\delta \bullet \vec{b}_\delta$. This process has the smallest rate because the broken symmetry occurs only once, through a single change in one derivative. This type of broken gauge symmetry is a form of particle-hole excitation/de-excitation, as opposed to a pure Umklapp process, in which the CM wave vector changes by a reciprocal lattice vector, defined by the original bulk lattice (as in Eq. 12c).

The most coherent excitation/de-excitation phenomena occur through pure Umklapp processes. These processes occur when a common set of derivatives change in the same way, at many points, as a result of symmetry. The possible differences in flux that result from Umklapp processes are defined by the ranges of possible values of the difference ($k_{cm} - k_{cm}'$) between the initial state ($k_{cm}$) and final state ($k_{cm}'$) CM wave vectors. By construction, for an Umklapp process to take place, this difference equals a reciprocal lattice vector, $\vec{G} = \sum_{\alpha=1,3} i_\alpha \vec{g}_\alpha$ ($i_\alpha$=integer), or $\vec{G} + \vec{k}$, where (as in Eq. 12b) $\vec{k}$ is restricted to the First Brillouin Zone. When a common set of derivatives change in the same way, particular contributions to $k_{cm} - k_{cm}'$ can be conserved along specific directions. When this occurs, one or more of these contributions (from a particular wave-vector or wave-vectors) can be repetitively added in the exponential factors (associated with Eq. 13) in the evaluation of particular matrix elements from flux contributions involving different Bloch states. Effectively, this repetitive addition of a particular contribution rigidly transfers momentum from all of the particles in the bulk into motion of the CM as a whole or through flux contributions at the boundaries into excited state (hole-like) coordinates. When the momentum is conserved in the initial and final states along two directions parallel to two primitive reciprocal lattice vectors $\vec{g}_\alpha$ and $\vec{g}_\delta$, each integer ($i_\alpha$ or $i_\delta$) associated with each contribution ($\frac{i_\alpha \vec{g}_\alpha}{2N_\alpha}$ or $\frac{i_\delta \vec{g}_\delta}{2N_\delta}$) to each wave-vector is conserved, which means that $k_{cm} - k_{cm}'$ vanishes identically along these directions for each value of $\frac{i_\beta \vec{g}_\beta}{2N_\beta}$ ($\beta \neq \alpha \neq \delta \neq \beta$). But then, in the summation in Eq. 26 over $i_\alpha$ and $i_\delta$, for each value of $i_\beta$, the contribution to the difference,

$$k_{cm} - k_{cm}' = \sum_{\{i_\eta \neq 0\}} \sum_{\eta=1,3} \frac{i_\eta \vec{g}_\eta}{2N_\eta}, \tag{29}$$

occurs only from values of $\frac{i_\eta \vec{g}_\eta}{2N_\eta} = \frac{i_\beta \vec{g}_\beta}{2N_\beta}$ that are parallel to $\vec{g}_\beta$. As a consequence, for each value of $i_\beta$, for all values of $i_\alpha$ and $i_\delta$ in the sum, a common contribution, $\frac{i_\beta \vec{g}_\beta}{2N_\beta}$, appears. But this means that the total contribution, $(k_{cm} - k_{cm}')|_{i_\beta}$, for a given value of $i_\beta$, is given by

$$(k_{cm} - k_{cm}')|_{i_\beta} = \sum_{\substack{i_\alpha=-N_\alpha, N_\alpha-1 \\ i_\delta=-N_\delta, N_\delta-1}} \frac{i_\beta \vec{g}_\beta}{2N_\beta} = 4N_\alpha N_\delta \frac{i_\beta \vec{g}_\beta}{2N_\beta}. \tag{30}$$

Also, in principle, in Eq. 29, the sum over non-vanishing values of $i_\eta$ (denoted by $\{i_\eta \neq 0\}$ in Eq. 29), $i_\eta = i_\beta$ can take on any value between $-N_\beta$ and $N_\beta - 1$, and the total number of terms in this sum can vary between 1 and $2N_\beta$. Since any number of the values of $i_\beta$ can be identical, it is convenient to designate each term in the associated set (from the set $\{i_\beta = i_\eta \neq 0\}$) with a separate index j; so that $i_\beta \equiv i_{\beta,j}$ is the j[th] element in the set. With these definitions, it follows that

$$(k_{cm} - k_{cm}') = \sum_j (k_{cm} - k_{cm}')|_{i_{\beta,j}} = 4N_\alpha N_\delta \frac{\vec{g}_\beta}{2N_\beta} \sum_j i_{\beta,j}. \tag{31}$$

A pure Umklapp process (without particle-hole excitation/de-excitation) occurs whenever the coefficient of $\vec{g}_\beta$ in Eq. 30 is an integer. This occurs identically when $N_\beta = i_\beta \times N_\alpha \times N_\delta; \quad \beta \neq \alpha \neq \delta \neq \beta$ (for all integer values of $i_\beta$ and $N_\beta$). When the coefficient of $\vec{g}_\beta$ is greater than one but not equal to an integer, in addition to a pure Umklapp process, the change in CM momentum includes particle-hole excitation/de-excitation. This occurs when $1 < \frac{2 \times i_\beta \times N_\alpha \times N_\delta}{N_\delta} \neq$ integer; $\beta \neq \alpha \neq \delta \neq \beta$.

As mentioned above, pure Umklapp processes can cause a large amount of momentum, effectively, to be transferred rigidly from all of the particles in the bulk through flux contributions at the boundaries of the bulk into excited state (hole-like) coordinates and into the motion of the CM, as a whole. The largest amount of momentum that can be transferred in this manner occurs when (as in Eqs. 30 and 31) the CM momentum is conserved along the directions that are parallel to $\vec{g}_\alpha$ and $\vec{g}_\delta$, and along the third direction, the magnitude of $(k_{cm} - k_{cm}')|_{i_{\beta,j}}$ is maximized. For this case, the same, common value $i_{\beta,j} = 2N_\beta - 1$ appears in each term, in the sum in Eq.31, and none of the terms, associated with any of the $2N_\beta$ (particle or hole) coordinates in the band, vanish. For this case, the contribution to each term in the sum is $4N_\alpha N_\delta \frac{(2N_\beta - 1)\vec{g}_\beta}{2N_\beta}$. Since the sum extends over $2N_\beta$ values of $i_{\beta,j} = 2N_\beta - 1$, the magnitude of the total CM momentum $\Delta P_{CM}^{max}$ is given by

$$|\Delta P_{CM}^{max}| = \hbar(8N_\alpha N_\delta N_\beta - 4N_\alpha N_\delta)|\vec{g}_\beta| = (N - N_{2d}^{\alpha\delta})\hbar|\vec{g}_\beta|, \tag{32}$$

where $N_{2d}^{\alpha\delta} = 4N_\alpha N_\delta$ is the number of unit cells in the two dimensional lattice, defined by the primitive vectors $\vec{b}_\alpha$ and $\vec{b}_\delta$.

Eq. 32 illustrates that the maximum amount of momentum that can be transferred to the lattice from the Umklapp process is size dependent and, for sufficiently large lattices, scales linearly with the total number ($N$) of unit cells in the lattice. In principle, this kind of Umklapp process can take place in any sub-lattice, contained in the original lattice, in which Eq. 30 is valid, which leads to the constraint that $N_\alpha \geq 1$, $N_\beta \geq 1$, $N_\alpha \geq 1$. In fact, less stringent constraints apply in situations involving reduced symmetry: When a finite (even) number of unit cells are periodically ordered along one or more directions (parallel to one or more of the primitive vectors) a pure Umklapp process can take place in which some maximum amount of momentum can be transferred to the CM of the sub-lattice. Details about this are discussed elsewhere ( Chubb 2005D ).

Because, in the presence of an applied field, the change in momentum that results from the semiclassical equations, increases linearly with time, and because Eq. 17 applies only in the limit of large values of $\tau$, to adequately address situations involving the application of applied fields, for extended periods of time, based on the semiclassical equations, it is necessary to incorporate the possibility that charge can accumulate outside the bulk region through (potentially time-dependent) variations in the zeroes of energy and momentum. In principle, the associated, non-unitary coupling to additional, resonant processes, can be included through a dependence in each many-body state on additional unoccupied states associated with higher energy band eigenvalues (that result from scattering processes). In this case, however, in order to accommodate the potential loss of gauge symmetry and non-unitarity, and the possibility that charge can accumulate, the associated wave-vectors are no-longer required to be real, or to be restricted to the First Brillouin Zone, the applicability of the formalism associated with low T processes (presented here), in general, need not apply, and a variant of the kind of heuristic approach ( Burcin Unlu et al 2004 ) that has been employed, involving semi-empirical Hamiltonians, would be required ( Chubb 2005D )

In the presence of applied fields, however, a net flux of particles (associated with each matrix element in Eq. 16) at the boundary of the bulk region occurs. Such a flux results in a "rigid" translation of the bulk that is required to break the gauge symmetry and degeneracy (associated with a perfectly rigid Umklapp process) because of the loss of translation symmetry. For this reason, in most situations, the kind of pure Umklapp process that could result in the maximum amount of momentum being transferred to the CM (as in Eq. 32, or in situations involving reduced symmetry [ Chubb 2005D ] ), in general, will not take place. However, when the highest occupied fermion band is completely filled, and the highest (lowest) occupied fermion (boson) band has negligible dispersion, when external forces are applied for appreciable periods of time, the kinds of coherent, rigid, Umklapp processes, associated with Eq. 32, can take place. In particular, provided the zero of energy and momentum associated with each hole-or particle- like coordinate remains fixed relative to the zero of energy and momentum of every other coordinate, no excitation or de-excitation of the bulk takes place. This can occur, rigorously, for all values of time t, provided that for all values of k,

$$\varepsilon(k)=\varepsilon(0)=\text{constant} \qquad . (33)$$

This can also occur over finite, discrete intervals (integer multiples of the period of a Bloch oscillation [ Bloch 1928 ] ), defined by the condition that

$$\int_{t_o}^{t_o+T^\alpha_{Bloch}} dt \frac{dk}{dt} \bullet \nabla_k \varepsilon_j (\frac{dk^\alpha}{dt}(t-t_o)+k_o) = \varepsilon_j(\vec{g}_\alpha+k_o)-\varepsilon_j(k_o)=0, \quad (34)$$

where $T^\alpha_{Bloch} = \frac{|\vec{g}_\alpha|}{|\frac{dk^\alpha}{dt}|} = \hbar \frac{|\vec{g}_\alpha|}{|e_j\vec{E}|}$ ($\vec{E}$ = constant, applied electric field, $e_j$ = charge of particle associated

with eignevalue $\varepsilon_j(k)$) is a particular, critical, interval of time (referred to as the Bloch oscillation period) that is associated with the potential breakdown of the single-band, semiclassical equations.

In the complete picture (presented here), these semiclassical equations only apply rigorously, in a perfect, finite lattice, to the GS and the lowest-lying excitations since they result from the relative shifts in the zero of momentum that result from broken gauge symmetry. (This is in contrast to the conventional quasi-particle picture where it is assumed they apply to states in higher energy bands, throughout an infinitely repeating lattice.) There is an additional restriction that naturally occurs in the finite lattice situation, (implicitly from Eq. 34), that is not obvious in the quasi-particle picture: Because the semiclassical equations result from broken gauge symmetry and apply to the GS, they do not include collisions and preserve time-reversal invariance. In particular, they only apply rigorously to the band that has the lowest energy unoccupied states. They apply only approximately in situations in which excitations into higher energy bands take place. For this reason, the potential breakdown of the single-band approximation in the conventional quasi-particle picture (for example, through Zener tunneling [ Zener 1934 ]), in general, involves a complicated many-body problem.

When Zener developed his tunneling arguments, using the semiclassical equations, implicitly, he introduced irreversible forms of scattering that are not rigorous. (Bloch [1928] used a single band picture, which actually is more rigorous because the associated Bloch Oscillations are assumed to be the result of elastic scattering processes, and this is necessary at sufficiently low T.) Implicitly the requirement that the semiclassical equations apply to a single band is equivalent to limiting the possible momentum ($\hbar \frac{dk}{dt}(t-t_o)$) that can be transferred by the constant external force $\hbar \frac{dk}{dt}$ over the interval of time $t-t_o$ to a particular set of indistinguishable particles, in the many body state. In solids, this is possible when the excitation processes do not involve coupling to phonons within the lattice. This constraint, in turn, can be imposed by requiring the magnitude of the change in energy between any two times, $t_1$ and $t_2$, not exceed the minimal energy ($\hbar \Omega_{phonon}^{min}$) that is required to excite phonons possessing a particular, minimal frequency ($\Omega_{phonon}^{min}$), defined by the lattice:

$$|\int_{t_1}^{t_2} dt \frac{dk}{dt} \bullet \nabla_k \varepsilon_j | = |\varepsilon_j(k(t_2)) - \varepsilon_j(k(t_1))| \leq \hbar \Omega_{phonon}^{min} \qquad (35)$$

Bloch (1928) recognized that in 1-dimension, the semiclassical equations predict that, in the context of the heuristic "quasi-particle" picture, in the absence of collisions, the effective electron momentum distribution function (defined by wave vectors in the First Brillouin Zone) could oscillate, as a function of time, indefinitely, with a period defined by $T^\alpha{}_{Bloch}$. Effectively, he also assumed that the theory would apply only to a single band.

In fact, the semiclassical equations do apply to a single band rigorously in finite but vanishing-ly small T. As a consequence, Bloch's use of the single band approximation is rigorous provided Eq. 35 is valid. However, in electron systems, this equation is never satisfied except in unusual situations. For this reason, Bloch oscillations do not take place in any "naturally-occurring" solid. Their existence has been inferred from optically induced excitations in artificially-formed multilayer, GaAs –AlGa hetero-structures ( Feldmann et al 1992 ) and observed directly through resonant tunneling of cold, neutral atoms between the two lowest bands in an optical lattice, by Dahan et al (1996) and Wilkinson et al ( 1996 ), and subsequently observed in absorption images of coherent matter wave, Bose Einstein Condensate (BEC) propagation in optical lattices by Morsch et al ( 2001 ), and Denschlag et al ( 2002 ). An intuitive explanation for the fact that the effect is difficult to produce exists: Bloch oscillation periods are so long that collisions with phonons and lattice imperfections always take place before a single Bloch oscillation period has elapsed. In the work by Feldman et al, involving GaAs –AlGa supercells, by construction, the energy bands along a particular direction can be forced to have negligible dispersion, and as a consequence, the inequality in Eq. 35 is satisfied.

An important reason that in the studies involving coherent matter waves in optical lattices ( Morsch et al 2001 , Denschlag et al 2002 ), the effect could be observed directly is that the associated lattice and "quasi-particles" are formed with such a high degree of control that the effects of collisions can be eliminated, and the applied external force can be finely tuned. In particular, especially in the work involving BEC's: 1.The lattice, which is formed from counter-propagating laser beams, does not have phonons and imperfections (except for those associated with boundary and finite size limitations); 2. There is no long-range Coulomb interaction; 3. When the BEC density is sufficiently dilute, the quasi-particle oscillations are essentially collision-less, and with increasing density, the effects of collisions can be monitored and controlled; and 4. As opposed to creating the external force from an applied $\vec{E}$ field, which is subject to non-uniformity, especially near surfaces and interfaces, in this work (Morsch et al 2001 , Denschlag et al 2002 ), the electrostatic force ($e_j \vec{E}$) (that appears in the conduction of charged particles in solids) is replaced by an effective force (defined by the relative acceleration of the lattice with respect to the CM of the BEC), which is: 1.Independent of the charge; and 2. Can be precisely controlled.  In particular, the velocity of the lattice can be changed arbitrarily by introducing an off-set in the frequency of one of the counter-propagating lasers.  When the off-set is a constant, the lattice moves at constant velocity, relative to the CM of the BEC.  When the off-set increases linearly with time, the lattice is uniformly accelerated, by a constant amount.

Since the changes in wave-vector (and transport) result from externally applied (inertial or gravitational), neutral forces, and the lattice is constructed using a dipole potential, formed from finely-tuned lasers, no coupling to phonons is required, and the upper limit in the inequality in Eq. 35, is defined either by the energy associated with the lifetime of the experiment, or by the energy band gap W (=the minimum difference in energy between the lowest energy band and the first excited state band),  not by $\hbar \Omega_{phonon}^{min}$. Because the force can be approximately uniform, provided the BEC is loaded into the lattice with an approximately uniform density (which can be accomplished using suitable Magneto-Optical-Trapping parameters [Denschlag et al 2002 ], and laser frequencies),  except in the immediate vicinity of the boundaries of the lattice, in the rest frame of the BEC, the net, effective force that acts on the CM of the BEC (as in the situation associated with CM motion of the bulk, in Eqs. 19a,b)  is directed opposite to and with the same magnitude as the force associated with the acceleration (provided by the lasers).   As a consequence, provided the BEC occupies one of the states in the lowest energy band, the semiclassical equations can be expected to apply rigorously, provided the time-scales associated with collisions are considerably longer than those that apply to the propagation of the BEC in the lattice.  In particular, this can be accomplished, for example, by adiabatically turning on the lattice potential ( Denschlag et al  2002 ) and by preparing the BEC with a density that is sufficiently diffuse that the Thomas Fermi limit (Dalfovo et al  1999 ) applies, and the time-scale $T_{TF} = \frac{\hbar}{\mu}$ ($\mu$=chemical potential) , associated with collision-induced fluctuations of the BEC, is considerably longer than the lifetime of the experiment.

The quasi-particle picture has limitations.   A more precise model (as presented here) is based the onset of broken gauge symmetry in finite lattices.  In particular, Eq. 33 is the true limit that describes a perfect insulator, in a finite lattice.  It also describes the limit associated with non-interacting, neutral atoms in an optical lattice. In the absence of magnetic fields, in a solid, $\hbar \frac{dk}{dt} = e_j \vec{E}$ for a particular charge $e_j$; in an optical lattice, the analogous equation is $\hbar \frac{dk}{dt} = Ma$, M=mass of neutral atom, a=acceleration of atom, relative to the optical lattice.  In either limit, the physical transport of particles can occur through a rigid translation of the CM of the lattice, as opposed to a form of tunneling.  When this translation is stifled for a long enough period of time, large amounts of momentum can be transferred, effectively, in a coherent manner, to the CM of the particles through matrix elements that involve different CM momenta (as in Eq. 32), associated with Umklapp processes.

Recently, it has been suggested ( Chubb, S.R.  2005B ;  2005C ) that precisely this alternative, coherent form of  "Zener breakdown," can occur through a process, in which charge that is responsible for the insulator-to-conductor transition comes from  (positively charged) ions of H (and its isotopes), as opposed to electrons.  When this coherent form of "Zener breakdown" involves $D^+$ ions (d's), the transition is from an insulating to a superconducting state since d's are bosons. In the limit in which a finite size crystal contains approximately one H or D atom per Pd atom, additional forcing of an H or D into the lattice can lead to these forms of ionic conduction in order to minimize energy ( Chubb, S.R. 2005B ; 2005C ). This can be expected to take place in finite size crystals, provided the deviations from full-loading (x=1 in $PdH_x$ or $PdD_x$) are sufficiently small because of the 4d-5s like, anti-bonding, orbital character and delocalized (Pd-like) behaviour of the electronic states ( Klein & Cohen 1992 ) immediately above and below the Fermi Energy ( Chubb, S.R. 2005A ).

An important distinction exists between cases involving finite, and infinitely repeating, periodic PdH and/or PdD lattices. In finite PdH and/or PdD crystal lattices, in order to sustain high-loading (x→1 in $PdH_x$ or $PdD_x$ ), externally applied electric (in the case of electrolysis) or pressure (in the case of gas-loading) fields are necessary, which means that even small variations in loading will induce variations in charge that extend throughout the crystal and into and away from the external regions that surround it ( Chubb, S.R. 2005A ; 2005B ). For this reason, in finite-size crystals, these fluctuations can carry small amounts of ionic charge away from the crystal. As a consequence, a finite size crystal, containing PdH or PdD, can conduct both electrons and protons (p's) or d's. This is not true in infinitely repeating PdH or PdD lattices, where the resulting coupling involves Pd-like electrons and the acoustical phonons.

The effect can occur in finite size crystals (but not in infinitely-repeating crystals) because the externally applied electric or pressure fields impart forces to the crystal. The effect can take place in PdH and PdD because (as a consequence of the electronic structure), provided the fluctuations in loading are small enough, the associated forces can be small enough that the associated many-body system can remain close to its GS. Effectively, this induces a shift in the zeroes of energy and momentum of the small ionic component associated with the fluctuation, relative to the comparable electronic component.

With increasing time, the amount of momentum increases. Each time the momentum exceeds a particular value associated with an Umklapp process, it becomes possible to alter the coordinates associated with holes and particles through an excitation of the system. But in the (effectively adiabatic) limit, in which the induced field (associated with the fluctuations) is sufficiently small, no excitation takes place, and effectively, the lattice is allowed to move rigidly. Eventually the momentum exceeds the maximal amount defined by the perfect, flat band limit, associated with Eq. (30). At this point, ion conduction takes place. In the limit of an infinitely repeating, periodic solid, the lattice becomes infinitely large and the externally applied field becomes infinitesimal. But in this limit, the time that is required for the onset of ion conduction becomes infinite.

Realistic estimates ( Chubb, S.R. 2005A; 2005B ), for a situation involving electrolytic loading, suggest that in finite size PdD crystals, effectively, even small fluctuations in loading ( $x = 1 \pm \delta$, $\delta \sim 3 \times 10^{-3}$, in $PdH_x$ or $PdD_x$ ) can induce changes in momentum that may be large enough to account for some of the anomalies that have been observed in these systems ( Arata & Zhang 1995 ; 1997A ; 1997B ; 1999 ; 2000A ; 2000B, Miles et al 2001 ). But whether or not this happens depends on the size of the crystal. In crystals that have characteristic dimensions of .01 millimeters, the oscillations can carry charge but only after a very long time (on the order of weeks). In crystals that have characteristic sizes of 10's of $\overset{o}{A}$, the applied electric fields can induce ion conduction in less than a hundredth of a second ( Chubb, S.R. 2005C ; 2005D ). The range of the possible Zener tunneling times (between fractions of a second and weeks), in these calculations, is consistent with the observed range of possible times associated with the onset of heat production, after the attainment of high-loading. This suggests that the observation that after high-loading, a variable "incubation", or "triggering" time period is required to elapse before the onset of heat, may be related to the sizes of the crystals that are used in the experiments. The fact that the calculations suggest that the shortest times occur in smaller crystals may explain why the anomalous heat appears to be more reliably reproduced in systems that have crystals with characteristic dimensions that are on the scale of 10's-100's of nm ( Arata & Zhang 1995 ; 1997A ; 1997B ; 1999 ; 2000A ; 2000B, Miles et al 2001 ). Arata and Zhang (2005), in particular, independently suggested a similar idea, based on evidence that they obtained in their experiments, involving Pd black.

In principle, the Generalized Multiple Scattering formalism can be used to estimate the rates associated with arbitrary forms of reactions, in finite regions of space. Thus, it can be used to estimate limiting time-scales, associated with finite-size effects. Implicitly, this fact has been used throughout this paper. For example, starting with Eq. 16, it is possible to estimate the time-scales associated with the impact of finite-size effects on dissipative processes involving collisions or with other effects associated with external forces that are neglected in existing protocols. These estimates are possible when it is possible to model the many-body wave functions of both an initial (unperturbed) and a final (perturbed) Hamiltonian, in which the perturbation $\Delta V$ is associated with a particular process or effect. For example, in modeling the effects of collisions, $\Delta V$ could be equated with the non-linear terms in the Gross-Pitaevskii equation (Dalfovo et al 1999). Similarly, $\Delta V$ could be equated with the change in effective potential energy associated with corrections to the initial Hamiltonian that can result when the relative acceleration ($\vec{a}$) of the BEC with respect to the lattice is non-uniform. For illustrative purposes, an estimate is given (immediately below) of the limiting time-scale associated with finite-size effects for this kind of situation. (The associated argument can also be applied to provide a comparable estimate of the limiting time-scale associated with finite size effects for a situation involving collisions [ Chubb 2005D ].)

When $\vec{a}$ is not uniformly constant over a particular region, but the BEC moves rigidly, to a first approximation, provided the associated non-uniformity in $\vec{a}$ varies slowly, the dependence of the many-body function $\Psi(r_1,...,r_{N_b})$ on the CM coordinates ($r_{cm}$) becomes separable from its dependence on each of the remaining, independent coordinates ($r_j - r_{cm}, j=1, N_b$). (Here, $N_b$ refers to the total number of bosons.) But then,

$$\Psi(r_1,...,r_{N_b}) = \Phi(r_{cm})\Psi_{rel}(r_1 - r_{cm},....,r_{N_b} - r_{cm}), \qquad (36)$$

where $\Phi(r_{cm})$ and $\Psi_{rel}(r_1 - r_{cm},....,r_{N_b} - r_{cm})$, respectively, describe the dependence of $\Psi(r_1,...,r_{N_b})$ on CM and the remaining (independent) coordinates, of the BEC; also, in this limit, at a particular location $r$, $\vec{a} = \vec{a}(r)$ can be expressed in terms of its linear order Taylor Series expansion relative to its value at an initial position ($r_o$) of the CM of the BEC.

Again, for illustrative purposes, it is convenient to consider a particular (idealized) situation, in which the BEC is in free fall (along the z-axis), the local gravitational constant g only varies in the z-direction, the local gravitational acceleration is

$$-g\hat{z} \equiv -g(z)\hat{z} = -(g(z_{cm}) + (z - z_{cm})\frac{\partial g(z_{cm})}{\partial z_{cm}})\hat{z}, \qquad (37)$$

and the lattice is also accelerated in the (negative) z-direction, uniformly, by an amount $|\vec{a}| = g(z_{cm})$. For this situation, in the frame of the lattice, the relative acceleration of the BEC is $\vec{a}_{latt}(z) = (-g(z) + g(z_{cm}))\hat{z}$, and the effective potential $V(z)$ that describes the CM motion is given by

$$V(z) = M_{BEC} \frac{(z-z_{cm})^2}{2} \frac{\partial g(z_{cm})}{\partial z_{cm}}. \qquad (38)$$

Through Eq. 16, it is possible to identify the minimum rate associated with the onset of finite size effects in measurements of $\left.\frac{\partial g(z_{cm})}{\partial z_{cm}}\right|_{z_{cm}=z_1} = g'(z_1)$ in a situation involving a particular value $z_{cm} = z_1$, relative to a comparable measurement of $\left.\frac{\partial g(z_{cm})}{\partial z_{cm}}\right|_{z_{cm}=z_2} = g'(z_2)$ that involves a different value of $z_{cm} = z_2$. By minimizing this rate, it is possible to identify critical values of the parameters (the number of lattice sites, $N$, and bosons, $N_B$) that are required to perform measurements of $\frac{\partial g(z_{cm})}{\partial z_{cm}}$ with a specified level of accuracy. In addressing this last problem, we can require that $\frac{\partial g(z_{cm})}{\partial z_{cm}}$ be non-negative. (When $\frac{\partial g(z_{cm})}{\partial z_{cm}} < 0$, the net contributions at the boundaries of the lattice to the rate expression, in Eq. 16, asymptotically vanish since the associated solutions of the Schroedinger equation, which involve Hermite polynomials with imaginary arguments, are oscillatory. As a consequence, in the absence of additional perturbations, the time-scale associated with the onset of finite size effects, for this case, is non-physical since it is always infinite.)

For a particular, important application (involving gravity gradient measurements from airplanes [van Leeuwen 2001]), a measurement of $\frac{\partial g(z_{cm})}{\partial z_{cm}}$ is required to have an absolute error of 1 Eotvos (=E =$10^{-9}$ m/(m-s) ~$10^{-10}$ g/m). It is possible to use this value to obtain approximate bounds for critical values of various parameters associated with the lattice and BEC (that result from the onset of finite size effects) by minimizing the Reaction rate (from Eq. 16) that results from the overlap of an initial state BEC, which is in an eigenstate of a Hamiltonian $H_o$, associated with an initial potential $V_o(z) = M_{BEC} \frac{(z-z_1)^2}{2} g'(z_1)$, defined by the mass (=$M_{BEC}$) of the BEC and its gradient ($g'(z_1)$) at z=$z_1$ with an eigenstate of a Hamiltonian $H_F$, which is identical to $H_o$, except that the Taylor series expansion is evaluated at the location z= $z_2$. The associated perturbation is

$$\Delta V(r) = V_F(r) - V_o(r) = M_{BEC} \left( \frac{(z-z_2)^2}{2} g'(z_2) - \frac{(z-z_1)^2}{2} g'(z_1) \right).$$

In a situation involving an infinite lattice, the initial and final states asymptotically approach the solutions of two, different, single-particle, harmonic oscillator Hamiltonians (that respectively have angular frequencies $\omega_1 = \sqrt{g'(z_1)}$ and $\omega_2 = \sqrt{g'(z_2)}$)). In the many-body situation, in a finite lattice, as a consequence of Eq. 6, minimal reaction rate occurs in the limit in which $\Psi_{rel}(r_1 - r_{cm}, ..., r_{N_b} - r_{cm})$ is the same in the initial and final states (so that the net flux from the relative coordinates vanishes), and the contribution from the CM coordinates, associated with the LS and RS of Eq. 6, is identically the same. This occurs when the CM portions of the initial and final state (which can both be assumed to be uniform and constant in directions perpendicular to the z-direction), respectively, can be described by wave functions $\Phi_o(r) = \dfrac{\phi_o(z)}{A^{1/2}}$ and $\Phi_f(r) = \dfrac{\phi_f(z)}{A^{1/2}}$ that are given by the GS wave functions defined by the cross-sectional area $A$ perpendicular to z, and the harmonic oscillator Hamiltonians, possessing frequencies $\omega_1$ and $\omega_2$:

$$\Phi_o(r) = \frac{1}{A^{1/2}} \left( \frac{M_{BEC}\omega_1}{\pi\hbar} \right)^{1/4} \exp\left(-\frac{M_{BEC}\omega_1 z^2}{2\hbar}\right), \tag{39a}$$

and

$$\Phi_f(r) = \frac{1}{A^{1/2}} \left( \frac{M_{BEC}\omega_2}{\pi\hbar} \right)^{1/4} \exp\left(-\frac{M_{BEC}\omega_2 z^2}{2\hbar}\right). \tag{39b}$$

It follows that when Eqs. 39a and 39b are substituted into Eq. 36, $\Psi_o = \Phi_o(r_{cm})\Psi_{rel}$ and $\Psi_f = \Phi_f(r_{cm})\Psi_{rel}$, and these wave functions can be used to evaluate the flux, when they are substituted into Eq. 6 (with $\Psi_f = \Psi'$, $\Psi_o = \Psi_{GS}$). Then, when $\Psi_{rel}(r_1 - r_{cm}, ..., r_{N_b} - r_{cm})$ is the same in the initial and final states (which is required for a minimal rate contribution from flux terms associated with the boundary), it follows that

$$\int d^3r \nabla \bullet <\Psi'|v(r)|\Psi_{GS}> = \int_{\partial V} dS\, \hat{n} \bullet <\Psi'|v(r)|\Psi_{GS}> = i\phi_o(L)\phi_f(L)2L(\omega_1 - \omega_2),$$

$$= i(\omega_1 - \omega_2)L e^{-\frac{(\omega_1+\omega_2)ML^2}{2\hbar}} (\omega_1\omega_2)^{1/4} \left( \frac{M_{BEC}}{\pi\hbar} \right)^{1/2}$$

(40a)

where $2L$ is the length of the lattice, and the factor of 2 (in the first line on the RS) is present because the contributions to the integral at each endpoint (defined by $r_{cm} = \pm L$) are equal. Also, consistent with the assumption that the contributions (from finite size effects) to the rate expression be minimized (in Eq. 17), the final density of states can be derived using the expression that applies to a single-particle, harmonic oscillator (with final frequency $\omega_2$):

$$\rho_{final}(E)\big|_{E=\frac{\hbar\omega_1}{2}} = \sum_F \delta(E - E_{exact}(C_F)) = \sum_n \delta\left(\frac{\hbar\omega_1}{2} - (n + \frac{1}{2})\hbar\omega_2\right) = \frac{1}{\hbar\omega_1}. \tag{40b}$$

It is possible to make an estimate of the minimal transition rate $R_{o \to f}$ (which can be used to infer critical values of the parameters associated with lattice site) between the initial and final states ($\Phi_o(r)$ and $\Phi_f(r)$) by substituting Eqs. 40a,b into Eq. 16:

$$R_{o \to f} = L^2(\omega_1\omega_2)^{1/2} \frac{M_{BEC}}{\hbar} \frac{(\omega_1 - \omega_2)^2}{\omega_1} e^{-\left(\frac{M_{BEC}(\omega_1+\omega_2)}{\hbar}L^2\right)}. \tag{41}$$

As a function of L, the maximum value of $R_{o \to f}$ can be used to infer a critical values ($L_{crit}$ and $2L_{crit}$). In particular, from the condition that $\left.\dfrac{\partial R_{o \to f}}{\partial L}\right|_{L=L_{crit}} = 0$, it follows that

$$L_{crit}^2 = \frac{\hbar}{M_{BEC}(\omega_1+\omega_2)}. \tag{42}$$

Since at the surface of the earth (Snadden et al 1998) $g'(z_1) \cong 3.08 \times 10^{-6} (Htz)^2 = 3080\,E \gg 1\,E = |g'(z_1) - g'(z_2)|$, with negligible error, the difference between $\omega_1$ and $\omega_2$ in the prefactor of $(\omega_1 - \omega_2)^2$ in Eq. 42 can be ignored. Then, it follows that the maximum value of $R_{o \to f}$ (=

$R_{o \to f}(L_{crit}) \cong \dfrac{(\omega_1 - \omega_2)^2}{\omega_1} e^{-1.0} \cong 2.010 \times 10^{-4}\,Htz$) does not depend on the value of $M_{BEC} = N_b M_{atom}$ and, thus, is independent of the number of bosons ($N_b$) and mass ($M_{atom}$) of an individual atom in the BEC. On the other hand, values of $L_{crit}$, lattice size ($=2 L_{crit} = 2N_z |b_z|$;

$$b_z = b_\alpha|_{\alpha=z} = \left.\frac{\bar{L}_\alpha}{2N_\alpha}\right|_{\alpha=z} ; \quad N_\alpha|_{\alpha=z} = N_z),$$ and $N_z$ all depend on $M_{BEC}$. In the initial measurements (

Anderson & Kasevich 1998) of g (based on optical lattices), the lattice spacing $|b_z|$=850nm, and (since, in this experiment, vapors of $^{87}Rb$ were used) $M_{atom} \cong 144.04 \times 10^{-24} g$. From these values, it follows that in this experiment, for $N_b = 1000000$, the minimal, required vertical length $L_{min} > 2l_{crit} \approx 912$ nm, and the minimal number of unit cells $N_{min} > 2$; while for $N_b = 1000$, $L_{min} > 2l_{crit} \approx 28833$ nm, $N_{min} > 68$, and for $N_b = 100$, $L_{min} > 2l_{crit} \approx 28833$ nm, $N_{min} > 214$.

# V Conclusions

In this paper, a new formalism, Generalized Multiple Scattering Theory, has been developed for estimating the rates of particular many-body processes, based on general properties of particular systems. This formalism is difficult to apply in many situations because it requires information about the fluxes of particles that enter and leave a particular region, where collisions are allowed to take place. The formalism is especially useful, however, for investigating the GS and lowest-energy excitations in situations involving broken gauge symmetry.

    In the case of ordered lattices, the relevant gauge symmetry involves the invariance of the many-body Hamiltonian with respect to Galilean transformations, in which the bulk region is translated rigidly without altering any of the particle-particle separations in the region. Broken gauge symmetry occurs through effects that result from accumulation of charge (in solids) or changes in the potentials (in optical lattices) at the boundaries of the periodically ordered (bulk) region. The requirement that, in the presence of a broken gauge symmetry, minimal overlap occur between the GS and the lowest-lying excited states imposes constraints on the form of the allowable states, in bulk regions. From this starting point, it is possible to prove a generalized form of Bloch's theorem and to generalize the associated semiclassical transport theory.

    Because Bloch's theorem, by construction, occurs as a result of a broken gauge symmetry, implicitly, it is associated with a (potentially huge) set of degenerate states. These states are related to each other through coherent forms of rigid-body motion, associated with Umklapp processes. Again, by construction, since this broken gauge symmetry becomes dominant when collisions are reduced, the associated Umklapp processes become important either at low T or when collisions are stifled. Quantitative bounds for the amounts of momentum that can be transferred from a crystal lattice to a surface or interface (and vice-versa) through these kinds processes, traditionally, in models in which the lattice is infinitely repeating and periodic, have been poorly defined. But in finite solids, at low, but finite T, precise, size-dependent bounds can be identified. In particular, although in larger crystals, collisions with phonons tend to reduce the magnitudes of the associated effects, in smaller crystals or in optical lattices, this is not the case.

    In a particular limit, in PdD and PdH crystals, a novel situation can occur in which $H^+$ and $D^+$ ions (in addition to electrons) can become charge carriers. At the transition between the ionic insulating- and conducting- state, a form of Zener/Ionic breakdown (similar to Zener/Electronic breakdown in insulators) in smaller PdD and PdH crystals can take place in which coherent Umklapp processes lead to effects that can be quite large. In the case of PdD, the associated effect can lead to a coherent form of interaction that mimics an insulator-

superconductor transition. The ranges of time-scales (which vary between weeks and fractions of a second) for initiating this kind of effect appear to be consistent with the comparable ranges of incubation times that have been observed to be required before anomalous heat is produced during the prolonged electrolysis of $D_2O$ by PdD. The fact that the shortest time-scales required for Zener/Ionic breakdown and the production of Excess Heat both occur in PdD crystals that have characteristic dimensions ~10's of nm suggests that Excess Heat is being triggered by Zener/Ionic breakdown.

Generalized Multiple Scattering Theory can be used to estimate the transition rate for a particular process, provided the process can be approximately represented, using known initial and final states. In the paper, a concrete example, involving the identification of critical length- and time- scales, associated with finite size, in the problem of measuring the gradient of the gravitational force, was used to illustrate this. This calculation, which illustrates that the critical length-scale of the lattice is inversely proportional to the square root of the number of atoms in a BEC, provides quantitative information that will be useful in the problem of observing variations in gravitational force, using coherent atom waves, from airborne platforms.

## Acknowledgements

I would like to acknowledge valuable discussions, which took place more than a decade ago, with Giuliano Preparata that inspired me to examine the relationship between broken gauge symmetry and electron transport in finite solids. I also would like to thank Talbot Chubb and David Nagel for their help and encouragement. The specific calculation, related to gravity gradient measurements, was partially inspired by computations performed through the DoD High Performance Computing program. The Office of Naval Research provided financial support for this work. Additional, partial support was provided through the advanced training program at the Naval Research Laboratory and through the Quantum Processes and Metrology group, in the Atomic Physics Division of the National Institute of Standards and Technology.